\documentstyle[art12]{article}

\def\magicphrase#1{{\it #1 can be 'embedded' into the ALH}}
\def\d{{\scriptstyle i \over \scriptstyle 2}\,[\delta]}

\begin{document}

\title{'Universality' of the Ablowitz-Ladik hierarchy.}
\author{V.E. Vekslerchik
\thanks{Regular Associate of the Abdus Salam International Centre for Theoretical Physics}
\\
\\
\normalsize\it
  Institute for Radiophysics and Electronics,
  Nat. Acad. of Sci. of Ukraine,
\\ \normalsize\it
  Proscura Street 12, Kharkov 310085, Ukraine.
\thanks{E-mail address: {\tt chik@ire.kharkov.ua}}
}
\date{\today}
\maketitle

\begin{abstract}
The aim of this paper is to summarize some recently obtained relations
between the Ablowitz-Ladik hierarchy (ALH) and other integrable equations.
It has been shown that solutions of finite subsystems of the ALH can be
used to derive a wide range of solutions for, e.g., the 2D Toda lattice,
nonlinear Schr\"odinger, Davey-Stewartson, Kadomtsev-Petviashvili (KP) and
some other equations.  Similar approach has been used to construct new
integrable models:  O(3,1) and multi-field sigma models.  Such
'universality' of the ALH becomes more transparent in the framework of the
Hirota's bilinear method. The ALH, which is usually considered as an
infinite set of differential-difference equations, has been presented as a
finite system of functional-difference equations, which can be viewed as a
generalization of the famous bilinear identities for the KP tau-functions.
\end{abstract}

\section{Introduction}
\setcounter{equation}{0}

One of the characteristic features of the theory of integrable systems is
the fact that various models, sometimes rather different apparently, turn
out to be, in such or other way, closely interrelated. For example, the
nonlinear Schr\"odinger equation (NLSE) is gauge equivalent to the
Landau-Lifshitz equation, sine-Gordon to Thirring model, Ablowitz-Ladik
model -- to the classical Heisenberg chain, etc. A large number of the
integrable equations can be obtained as reductions of, e.g., the KP or the
self-dual Yang-Mills equations. Poles of the rational solutions of the KdV
evolve according to the Calogero model and so on.

Another kind of interrelations has been discovered by Levi, Benguria
\cite{BL,L}, Flaschka \cite{F}, and also Shabat, Yamilov \cite{SY}.  They
considered sequences of Backlund transformations (BT) for some integrable
nonlinear problems

$$
N[\psi] = 0,
\qquad\qquad
... \to \psi \to \psi_{1} \to \psi_{2} \to ...
$$
and demonstrated that these sequences can be described by
differential-difference equations (DDE) which are also integrable. For
example, sequences of BT's for the NLSE provide solutions for the Toda
chain.

Similar ideas, in a somewhat transformed form, have been used in the works
\cite{sigma,2dtl,ds}. It turns out that
in some situations solutions of the DDE's can be used to generate families
of solutions for some related partial differential equations (PDE).
Namely such cases have been discussed in \cite{sigma,2dtl,ds}. There the
approach of \cite{BL,L,SY} has been extended by taking into account
several DDE's simultaneously, which enabled to deal with multidimensional
PDE's, such as the 2d Toda lattice (2DTL), and the Davey-Stewartson (DS)
equation (2+1-dimensional systems) and others. The method of the works
\cite{sigma,2dtl,ds} can be explained by discussing the following very
simple examples.

\bigskip
\noindent
{\it Toy example 1.}

\noindent
Consider the system of differential-difference equations

\begin{equation}
\left\{
\begin{array}{l}
i \partial_{x} q_{n} = q_{n+1} + q_{n-1}
\cr
\phantom{i} \partial_{y} q_{n} = q_{n+1} - q_{n-1}
\end{array}
\right.
\label{toy1}
\end{equation}
By trivial algebra one can show that, first, this system is compatible
and, second, that for every $n$ the quantity $q=q_{n}$
satisfies the Helmholtz equation.

\begin{equation}
\Delta q + 4 q = 0,
\qquad\qquad
\Delta=\partial_{xx}+\partial_{yy}
\label{Helmholtz}
\end{equation}
This example is aimed to demonstrate that the system of DDE's can be
converted to PDE. In the linear case this is not very interesting, but
in the nonlinear one, as will be shown below, such transformations may be
fruitful.

\bigskip
\noindent
{\it Toy example 2.}

\noindent
As the second example consider the system (\ref{toy1}) enlarged with
one equation more:

\begin{equation}
\left\{
\begin{array}{l}
i \partial_{x} q_{n} = q_{n+1} + q_{n-1}
\cr
\phantom{i} \partial_{y} q_{n} = q_{n+1} - q_{n-1}
\cr
i \partial_{t} q_{n} = q_{n+2} + q_{n-2}
\end{array}
\right.
\label{toy2}
\end{equation}
Again, one can straightforwardly verify that this system is
compatible and obtain that for every $n$ the quantity $q=q_{n}$ satisfies
the equation

\begin{equation}
i \partial_{t} q +
{1 \over 2} \Box q = 0,
\qquad\qquad
\Box=\partial_{xx}-\partial_{yy}
\end{equation}
As in the first example we have converted the system of DDE's into a PDE.
But it should be noted that we started with three 1+1, i.e. 2 dimensional
equations, while the resulting one is 2+1, i.e. 3 dimensional.  This is
not crucial in the linear case, but is very important in the nonlinear
one.  The methods elaborated for multidimensional ($d \ge 3$) nonlinear
integrable systems (2DTL, DS, KP, etc.) such as, e.g., the multidimensional
inverse scattering transform (MIST), are much more complicated than the
traditional one (IST) developed for the 2- and (1+1)-dimensional systems
(KdV, NLSE, Ablowitz-Ladik model, etc.).
It will be shown below that in the nonlinear case such 'decomposing'
of a multidimensional PDE in a system of low-dimensional DDE's
can be rather useful, at least from the practical viewpoint.

\bigskip

Transformations similar to the ones described above,
their nonlinear variant, are the core of the
works \cite{sigma,2dtl,ds}. But before proceeding further I would like to
discuss the following very important question, namely the question of
compatibility. It is rather easy to write down compatible linear systems
as the ones above, but in the nonlinear case the problem is a little bit
more difficult.  Nevertheless it is surely possible to construct systems,
say, ones similar to (\ref{toy1}), (\ref{toy2}). But there is no need to
do that, because such nonlinear and compatible systems are already known
-- they are the integrable hierarchies.  Every integrable equation does
not appear alone, it is always a member of some infinite set of related
equations.  And commutativity of corresponding flows, or involutivity of
corresponding Hamiltonians, or, in simpler words, compatibility of
corresponding equations, is one of the ingredients of integrability.
Thus, in what follows I will 'nonlinearize' the toy examples using an
integrable hierarchy, namely the Ablowitz-Ladik hierarchy.

The plan of the present paper is as follows. After outlining some basic
facts related to the ALH (section \ref{sec-ALH}) I will establish
relations between this hierarchy and some other equations, such as 2DTL,
DS, NLSE, KP (section \ref{sec-embedding}) as well as few vector models
(section \ref{sec-gauged}).  To make this 'universality' of the ALH
more transparent I will discuss it in the framework of the Hirota's
bilinear method and present in section \ref{sec-newell} the recently
obtained \cite{newell} functional representation of the ALH which can be
viewed as a generalization of the famous bilinear identities for the KP
tau-functions.

\section{Ablowitz-Ladik hierarchy.} \label{sec-ALH}
\setcounter{equation}{0}

The ALH is an infinite
set of ordinal differential-difference equations, that has been introduced
by Ablowitz and Ladik in 1975 \cite{AL1}. The most well known of these
equations is the discrete nonlinear Schr\"odinger equation (DNLSE)

\begin{equation}
i \dot q_{n} =
q_{n+1} - 2q_{n} + q_{n-1} -
q_{n}r_{n} \left( q_{n+1} + q_{n-1} \right)
\label{dnlse}
\end{equation}
and the discrete modified KdV equation (DMKdV) (see, e.g., \cite{AS}),

\begin{equation}
\dot q_{n} =
p_{n} \left( q_{n+1} - q_{n-1} \right)
\label{dmkdv}
\end{equation}
where
\begin{equation}
p_{n} = 1 - q_{n}r_{n}, \quad
r_{n} = -\kappa \bar q_{n}, \quad
\kappa = \pm 1
\label{p_n}
\end{equation}
All equations of the ALH can be presented as the compatibility condition
for the linear system

\begin{eqnarray}
\Psi_{n+1} &=& U_{n} \Psi_{n}
\label{zcr-sp}\\
\partial_{t} \Psi_{n} &=& V_{n} \Psi_{n}
\label{zcr-evol}
\end{eqnarray}
where $\partial_{t}$ stands for $\partial/\partial t$,
which leads to their zero-curvature representation:

\begin{equation}
\partial_{t} U_{n} = V_{n+1} U_{n} - U_{n} V_{n}
\label{ZCR}
\end{equation}
In the standard IST approach developed in \cite{AL1} the matrix $U_{n}$
for the ALH is given by

\begin{equation}
U_{n} =
\pmatrix{ \lambda & r_{n} \cr q_{n} & \lambda^{-1} }
\end{equation}
where $\lambda$ is the auxiliary (spectral) constant parameter.

According to \cite{AL1}, elements of the matrix $V_{n}$, can be chosen as
Laurent polynomials in $\lambda$ in such a way that (\ref{ZCR}) holds
automatically for all $\lambda$'s provided $q_{n}$'s and $r_{n}$'s satisfy
some differential relations.  It should be noted that one can obtain an
infinite number of the matrices $V_{n}$ (which are Laurent polynomials of
different order) which leads to the infinite number of differential
equations $\partial q_{n} / \partial t = F^{l}_{n}$, ($l=1,2,...$).
Using the widely accepted viewpoint one can consider $q_{n}$'s
and $r_{n}$'s as depending on the infinite number of 'times', $q_{n} =
q_{n}(t_{1}, t_{2}, ...)$ and consider the $l$th equation of the ALH as
describing the flow with respect to the $l$th variable, $\partial q_{n} /
\partial t_{l} = F^{l}_{n}$.
Traditionally it is implied that all 'times' $t_{l}$ are real,
which is grounded from the standpoint of physical applications, and also
is convenient in the framework of the inverse scattering technique.
However, in some cases it is more convenient to use
instead of real 'times' $t_{l}$ some complex variables $z_{j},
\bar z_{j}, \; j=1,2,...$ (as in, say, 2D Toda theory),
which, as will be shown below, exhibit in a more
explicit way some intrinsic properties of the ALH.  A simple analysis
yields that the family of possible solutions of
(\ref{ZCR}) (and hence the equations of the hierarchy) can
be divided in two subsystems.  One of them consists of $V$-matrices which
are polynomials in $\lambda^{-1}$ (I will term the corresponding equations
as a 'positive' part of hierarchy), and the other consists of matrices
which are polynomials in $\lambda$ ('negative' subhierarchy), while in the
standard, 'real-time', approach all $V$-matrices contain terms
proportional to $\lambda^{m}$ together with the terms proportional to
$\lambda^{-m}$ ($m \geq 0$).  Let us consider first the 'positive' case.
An infinite number of polynomial in $1/\lambda$ solutions $V_{n}^{j}$
($j=1,2,...$) possesses the following structure:

\begin{equation}
V_{n}^{j} = \lambda^{-2} V_{n}^{j-1} +
\pmatrix{ \lambda^{-2} \alpha_{n}^{j} & \lambda^{-1} \beta_{n}^{j} \cr
          \lambda^{-1} \gamma_{n}^{j} & \delta_{n}^{j} }
\end{equation}
where the elements $\alpha_{n}^{j}, ... , \delta_{n}^{j}$ satisfy the
equations

\begin{eqnarray}
&&
\alpha_{n+1}^{j} - \alpha_{n}^{j} =
   - q_{n} \beta_{n+1}^{j} + r_{n} \gamma_{n}^{j}
\\&&
\delta_{n+1}^{j} - \delta_{n}^{j} =
   q_{n} \beta_{n}^{j} - r_{n} \gamma_{n+1}^{j}
\\&&
\partial_{j} q_{n} =
  q_{n} \delta_{n+1}^{j} + \gamma_{n+1}^{j} =
  q_{n} \alpha_{n}^{j+1} + \gamma_{n}^{j+1}
\\&&
\partial_{j} r_{n} =
  - r_{n} \delta _{n}^{j} - \beta_{n}^{j} =
  - r_{n} \alpha_{n+1}^{j+1} - \beta_{n+1}^{j+1}
\end{eqnarray}
with $\partial_{j} = \partial / \partial z_{j}$.
Choosing

\begin{equation}
a_{n}^{0} = b_{n}^{0} = c_{n}^{0} = 0,
\qquad
d_{n}^{0} = - i
\end{equation}
we can obtain consequently

\begin{equation}
\begin{array}{l}
\alpha_{n}^{1} = 0
\cr
\beta_{n}^{1} = -i r_{n-1}
\cr
\gamma_{n}^{1} = -i q_{n}
\cr
\delta_{n}^{1} = i r_{n-1}q_{n}
\end{array}
\qquad
\begin{array}{l}
\alpha_{n}^{2} = - i r_{n-1}q_{n}
\cr
\beta_{n}^{2} = - i r_{n-2}p_{n-1} + i r^{2}_{n-1}q_{n}
\cr
\gamma_{n}^{2} = - i p_{n}q_{n+1} + i r_{n-1}q^{2}_{n}
\cr
\delta_{n}^{2} =
  i r_{n-2}p_{n-1}q_{n} + i r_{n-1}p_{n}q_{n+1} - i r^{2}_{n-1}q^{2}_{n}
\end{array}
\end{equation}
and, in principle, all other matrices $V_{n}^{j}$. This leads to the
infinite system of equations for $q_{n}$, $r_{n}$, some first of which are

\begin{eqnarray}
\partial_{1} q_{n} &=& -ip_{n}q_{n+1}
\label{1q}
\\
\partial_{1} r_{n} &=& \phantom{-} ir_{n-1}p_{n}
\label{1r}
\\
\cr
\partial_{2} q_{n} &=&
 -ip_{n}p_{n+1}q_{n+2}+ir_{n-1}p_{n}q_{n}q_{n+1}+ip_{n}r_{n}q_{n+1}^{2}
\label{2q}
\\
\partial_{2} r_{n} &=&
 \phantom{-}
 ir_{n-2}p_{n-1}p_{n}-ir_{n-1}p_{n}r_{n}q_{n+1}-ir_{n-1}^{2}p_{n}q_{n}
\label{2r}
\end{eqnarray}
Analogously, looking for the $V$-matrices of the form

\begin{equation}
V_{n}^{-j} = \lambda^{2} V_{n}^{-j+1} +
\pmatrix{ \alpha_{n}^{-j} & \lambda \beta_{n}^{-j} \cr
          \lambda \gamma_{n}^{-j} & \lambda^{2} \delta_{n}^{-j} }
\end{equation}
one can obtain the 'negative' part of the ALH. Some first of its equations
are

\begin{eqnarray}
\bar\partial_{1} q_{n} &=& -iq_{n-1}p_{n}
\label{-1q}
\\
\bar\partial_{1} r_{n} &=& \phantom{-} i p_{n}r_{n+1}
\\
\cr
\bar\partial_{2} q_{n} &=&
-iq_{n-2}p_{n-1}p_{n}+iq_{n-1}p_{n}q_{n}r_{n+1}+iq_{n-1}^{2}p_{n}r_{n}
\label{-2q}
\\
\bar\partial_{2} r_{n} &=&
 \phantom{-}
 ip_{n}p_{n+1}r_{n+2}-iq_{n-1}p_{n}r_{n}r_{n+1}-ip_{n}q_{n}r_{n+1}^{2}
\label{-2r}
\end{eqnarray}
where $\bar\partial_{j} = \partial_{-j} = \partial / \partial \bar z_{j}$
and $\bar z_{j}$ is the complex conjugated of $z_{j}$.

I will not discuss here this hierarchy in detail, because now we have all
the necessary to derive relations between the ALH and other integrable
models and namely this is the main theme of the present paper.

\section{'Embedding' into the ALH.} \label{sec-embedding}
\setcounter{equation}{0}

\subsection{The Ablowitz-Ladik hierarchy and the O(3,1) $\sigma$-model.}

The first, and the simplest, implementation of the
'embedding' into the ALH method was carried out in \cite{sigma} and can be
viewed as a nonlinearized version of the first toy
example.

Consider the following system of two equations of the ALH:

\begin{equation}
\left\{
\begin{array}{l}
i \partial_{x} q_{n} = p_{n}\left(q_{n+1} + q_{n-1} \right)
\cr
\phantom{i}
\partial_{y} q_{n} =  p_{n} \left( q_{n+1} - q_{n-1} \right)
\end{array}
\right.
\label{sigma-system}
\end{equation}
The first of them is the DNLSE (\ref{dnlse}) transformed by means of the
substitution $q_{n} \rightarrow q_{n}\exp(2ix)$, while the second is the
DMKdV (\ref{dmkdv}). One can also view these equations as (\ref{1q}),
(\ref{-1q}) rewritten in terms of the variables $x = {\rm Re}\, z_{1}$ and
$y = {\rm Im}\, z_{1}$.

This system is compatible because it is a system of equations belonging to
one hierarchy (this fact is not hard to verify directly in this case).
Differentiating the first equation with respect to $x$ and the second one
with respect to $y$ one can get the identity

\begin{equation}
{\rm div}\, {1 \over p_{n} } \nabla q_{n} +
2\left( p_{n-1} + p_{n+1} \right) q_{n} = 0
\end{equation}
from which, using again (\ref{sigma-system}), one can derive the
following one:

\begin{equation}
{\rm div}\, {1 \over p_{n} } \nabla q_{n} +
\left[ 4 -
{1 \over p_{n}^{2}} \left| \nabla q_{n} \right|^{2}
\right] q_{n} = 0
\label{sigma_n}
\end{equation}
Thus, we have obtained that for all $n$'s the quantities $q=q_{n}$
solve the PDE

\begin{equation}
\Delta q
- \kappa {\bar q \over p} \left( \nabla q, \nabla q \right)
+ 4pq = 0
\label{nonlin-Helmholtz}
\end{equation}
where

\begin{equation}
p = 1 + \kappa |q|^2,
\qquad
\Delta = \partial_{xx} + \partial_{yy}
\end{equation}
which is a nonlinear analog of the Helmholtz equation (\ref{Helmholtz}).
It turns out that this equation is not merely some 'nonlinearization' of
the Helmholtz equation. It has a rather
interesting physical origin. This is the field equation of some
$\sigma$-model which has been proposed and discussed in \cite{sigma} and
which  should be called, using the terminology by Pohlmeyer \cite{Pohl},
the O(3,1) $\sigma$-model.
This model arises from the problem

\begin{equation}
\Delta \phi^{\mu} + \lambda \phi^{\mu} = 0
\label{begin}
\end{equation}
under the restriction

\begin{equation}
\phi_{\mu}\phi^{\mu} = -1
\label{restr}
\end{equation}
Here $\Delta$ is the two-dimensional Laplacian, $\phi^{\mu}$ is a
space-like vector from the Minkowski space,

\begin{equation}
\phi^{\mu} = \left( \phi^{0}, \phi^{1}, \phi^{2}, \phi^{3} \right)
\end{equation}
with the scalar product

\begin{equation}
\phi_{\mu}\psi^{\mu} =
\phi^{0}\psi^{0} - \phi^{1}\psi^{1} - \phi^{2}\psi^{2} - \phi^{3}\psi^{3}
\end{equation}
To satisfy the condition (\ref{restr}) the Lagrange multiplier
$\lambda$ is to be set to
$\lambda = - \left( \nabla \phi_{\mu}, \nabla \phi^{\mu} \right)$.
The vectors $\phi^{\mu},\, \partial_{x}\phi^{\mu},\,
\partial_{y}\phi^{\mu}$ together with the time-like unit vector
$\chi^{\mu}$ which is normal to the surface $\phi^{\mu}(x,y)$,
form a local basis in the Minkowski space which satisfies the
Gauss-Weingarten system,

\begin{equation}
\partial_{x} {\cal F} = U {\cal F},
\qquad
\partial_{y} {\cal F} = V {\cal F},
\qquad
{\cal F} =
  \left(
  \phi^{\mu} ,\;
  \partial_{x}\phi^{\mu} ,\;
  \partial_{y}\phi^{\mu} ,\;
  \chi^{\mu}
  \right)^{T}
\label{GW}
\end{equation}
where $U$ and $V$ are some $4 \times 4$-matrices (not written here).
The integrability conditions for the system (\ref{GW}), the so-called
Gauss-Codazzi equations, can be presented as follows (see \cite{sigma} for
details):

\begin{equation}
\Delta \alpha - 4 \sin\alpha \cos\alpha +
{\cos\alpha \over \sin^{3}\alpha}
\left( \nabla\beta, \nabla\beta \right) = 0
\label{chiral1}
\end{equation}
and

\begin{equation}
{\rm div}\, \left( \cot^{2}\alpha \, \nabla\beta \right) = 0
\label{chiral2}
\end{equation}
(note that the particular case of the above equations, $\beta = 0$, is
the well-known elliptic sine-Gordon equation).
These equations are the Euler-Lagrange equations for the action

\begin{equation}
S = \int \!\!\! \int  dx dy \; {\cal L}
\label{S}
\end{equation}
with the Lagrangian

\begin{equation}
{\cal L} =  \left( \nabla\alpha , \nabla\alpha \right) +
\cot^{2}\alpha \left( \nabla\beta , \nabla\beta \right) -
4 \cos^{2}\alpha
\label{LA}
\end{equation}
which can be rewritten in terms of the function

\begin{equation}
q = \cos\alpha \cdot \exp(i \beta )
\end{equation}
as

\begin{equation}
{\cal L} =
{\left( \nabla q , \nabla \bar q \right) \over 1 - |q|^{2} } -
4 |q|^{2}
\label{L}\end{equation}

This Lagrangian can also be obtained as the reduction of the Euclidean
version of the principal chiral field model, as is outlined in the
Appendix of \cite{sigma}.

The field equation corresponding to the Lagrangian (\ref{L}),

\begin{equation}
\Delta q +
{\bar q \over 1 - |q|^{2}} \left( \nabla q, \nabla q \right) +
4\left(1 - |q|^{2}\right)q = 0
\label{laplas}
\end{equation}
is nothing other than equation (\ref{nonlin-Helmholtz}) for $\kappa=-1$.

The field equation (\ref{laplas}) is integrable and it is possible to
develop the inverse scattering scheme applicable to this equation (it has
been done in the paper \cite{sigma-ist}).  However, it turns out that the
derived relationship between this model and the ALH provides almost all
results that are usually obtained in the framework of the ISM.  It can be
used to demonstrate that the $O(3,1)$ $\sigma$-model possesses an infinite
number of symmetries and conserved quantities and to obtain soliton and
some other solutions for the field equation.

\subsection{The Ablowitz-Ladik hierarchy and the 2DTL.}

In this section I want to present some results that have no nontrivial
linear analogues. Let us consider again the DNLSE-DMKdV system, now
written in terms of the complex variables $z_{1}$ and $\bar z_{1}$, i.e.
the system (\ref{1q}), (\ref{-1q}):

\begin{equation}
\left\{
\begin{array}{l}
\partial     q_{n} = -i p_{n} q_{n+1}
\cr
\bar\partial q_{n} = -i p_{n} q_{n-1}
\end{array}
\right.
\label{toda-system}
\end{equation}
(here $\partial$ stands for $\partial / \partial z_{1}$
and $\bar\partial$ for $\partial / \partial \bar z_{1}$)
and turn our attention to the quantity $p_{n}$.  One can
derive from (\ref{toda-system}) the corresponding equations for $p_{n}$

\begin{equation}
\left\{
\begin{array}{l}
\partial     p_{n} =
  i\kappa p_{n} \left( \bar q_{n-1} q_{n} - \bar q_{n} q_{n+1} \right)
\cr
\bar\partial p_{n} =
  i\kappa p_{n} \left( q_{n} \bar q_{n+1} - q_{n-1} \bar q_{n} \right)
\end{array}
\right.
\label{p-system}
\end{equation}
and obtain from this system, together with (\ref{toda-system}), that the
quantities $p_{n}$ satisfy the following equation:

\begin{equation}
\partial\bar\partial \ln p_{n} = p_{n-1} - 2 p_{n} + p_{n+1}
\end{equation}
This equation is the famous 2DTL equation which can be rewritten  in
terms of the functions $u_{n}$ defined by
$ p_{n} = \exp \left( u_{n} - u_{n+1} \right) $ and the real variables
$x$ and $y$ ($z_{1} = x+iy$) as follows:

\begin{equation}
{ 1 \over 4 } \Delta u_{n} =
\exp \left( u_{n-1} - u_{n} \right) -
\exp \left( u_{n} - u_{n+1} \right)
\label{2DTL}
\end{equation}
Thus, we have shown that the 2D Toda lattice turns out to be
hidden in the simplest equations of the ALH or, in other terms,
\magicphrase{the 2DTL}.

The main, so to say, 'practical' value of the ALH--2DTL correspondence is
in the fact that we can obtain a wide number of results for the 2DTL,
which is 2+1 dimensional system, by means of the traditional version of
the inverse scattering transform which has been developed for the 1+1
dimensional systems such as DNLSE, DMKdV, without invoking the MIST (that
has been elaborated for the 2DTL by Lipovsky~V.D.  and Shirokov~A.V. in
\cite{LS1}).  All known solutions for the ALH (read for the DNLSE or DMKdV)
can be converted to solutions for the 2DTL.

Another interesting problem is the question of the conservation laws
\cite{KS}.  One can now obtain an infinite number of the conservation laws
in terms of the ALH and then 'convert' them to the 2DTL case (this idea
has been described in \cite{SY}). Some new (comparing with the paper
\cite{KS}) conservation laws have been obtained in \cite{2dtl} using the
following simple procedure. Take one of them

\begin{equation}
\mbox{div}\; \vec J = 0
\end{equation}
and apply operator $\partial / \partial t_{j}$ where $t_{j}$ is one of the
hierarchy's times. This will yield an infinite number of conservation laws

\begin{equation}
\mbox{div}\; \vec J_{j} = 0
\end{equation}
An interesting manifestation of the ALH-2DTL relation is the fact that
the so-called conserved quantities 'of discrete variable direction' which
are of the form

\begin{equation}
\partial I = 0
\end{equation}
derived by Kajiwara and Satsuma \cite{KS} when considered in the framework
of the approach of \cite{2dtl} are nothing other than constants of motion
of the equations of the ALH, i.e. Hamiltonians of the ALH-flows.

\subsection{The Ablowitz-Ladik hierarchy and the DS equation.}

Now let us discuss the bigger subsystems of the ALH, and start with the
nonlinearization of the second toy example. Consider the system
(\ref{1q}) - (\ref{-2q})

\begin{eqnarray}
\partial_{1} q_{n} &=& -ip_{n}q_{n+1}
\label{ds:1}
\\
\bar\partial_{1} q_{n} &=& -iq_{n-1}p_{n}
\label{ds:-1}
\\
\cr
\partial_{2} q_{n} &=&
 -ip_{n}p_{n+1}q_{n+2}
 +ir_{n-1}p_{n}q_{n}q_{n+1}
 +ip_{n}r_{n}q_{n+1}^{2}
\label{ds:2}
\\
\bar\partial_{2} q_{n} &=&
 -iq_{n-2}p_{n-1}p_{n}
 +iq_{n-1}p_{n}q_{n}r_{n+1}
 +iq_{n-1}^{2}p_{n}r_{n}
\label{ds:-2}
\end{eqnarray}
Here I do not write the corresponding equations for $r_{n}$'s, which can be
restored from (\ref{ds:1}) - (\ref{ds:-2}) using the involution
$r_{n} = - \kappa\bar q_{n}$.

Differentiating (\ref{ds:1}) with respect to $z_{1}$ one can
straightforwardly obtain, using (\ref{ds:2}), that

\begin{equation}
i \partial_{2} q_{n} + \partial_{1}^{2} q_{n} =
- 2 C_{n} q_{n}
\label{compds1}
\end{equation}
where $ C_{n} = r_{n-1} p_{n} q_{n+1}$. On the other hand, it follows
from (\ref{ds:-1}) that

\begin{equation}
\bar\partial_{1} C_{n} =
i p_{n} \left( r_{n}q_{n+1} - r_{n-1}q_{n} \right) =
\partial_{1} p_{n}
\label{compds2}
\end{equation}
which leads to

\begin{equation}
\Delta C_{n} =
4 \partial_{1}^{2} p_{n}
\end{equation}
where
$\Delta = 4 \partial_{1} \bar \partial_{1}$.
Performing analogous computations starting
from the equations (\ref{ds:-1}) and (\ref{ds:-2}) one can obtain similar
relations for $r_{n}$ and $\overline{C}_{n}=q_{n-1}p_{n}r_{n+1}$.
Using the real variables $x, y$ and $t$

\begin{equation}
x = {\rm Re}\, z_{1}, \qquad
y = {\rm Im}\, z_{1}, \qquad
t = {\rm Re}\, z_{2}
\label{xyt}
\end{equation}
and introducing the real quantities $A_{n}$ given by
$ A_{n} = {\rm Re}\, C_{n}$,

\begin{equation}
A_{n} =
{1 \over 2} p_{n}
\left( r_{n-1}q_{n+1} + q_{n-1}r_{n+1} \right),
\label{An}
\end{equation}
one can show that $q_{n}$ and $A_{n}$ satisfy the following system

\begin{eqnarray}
&&
i\partial_{t} q_{n} + {1 \over 2} \Box q_{n}  + 4 A_{n} q_{n} = 0
\label{dsq}\\ &&
\Delta A_{n} = \kappa \Box \left| q_{n} \right|^{2}
\label{dsa}
\end{eqnarray}
where $\Box = \partial_{xx} - \partial_{yy} $.  This system is nothing
other than the DS system \cite{DS} or, to be more precise, the DS-II
equation, according to the generally accepted classification.

In such a way we have obtained the following result: solutions of the
equations (\ref{ds:1}) -- (\ref{ds:-2}), belonging to the ALH, can be used
to obtain solutions for the DS equation, i.e. \magicphrase{the DS
equation}.  Moreover, when solving (\ref{ds:1}) -- (\ref{ds:-2}), we
obtain simultaneously an infinite (for an infinite ALH chain) number of
solutions for the DS: the pairs $(q_{n}, A_{n})$ for all $n$'s solve the
DS system.  Thus, equations (\ref{ds:1}) -- (\ref{ds:-2}) can be viewed as
sequences of the Backlund transformations.  And again, as in the case of
the 2DTL, we can obtain a wide number of solutions for this
(2+1)-dimensional system by, so to say, little efforts: without invoking
the MIST. But here we come to an important for this approach question.
Each solution of the ALH subsystem was shown to satisfy the DS. The
question is whether the reverse statement is true.  The answer is
negative.  We have seen that all $q_{n}$'s satisfy the equation of the
O(3,1) $\sigma$-model.  So, solutions that we obtain by 'embedding' into
the ALH are only some subclass of all possible solutions of the DS. But it
turns out that this subclass is rather rich: it contains solitons,
'Wronskian' solutions, finite-gap quasiperiodic ones (these solutions are
discussed in \cite{ds}) and many other. It seems to be interesting that
the pioneering papers by Ablowitz and Ladik \cite{AL1,AL2}, which have
been written twenty years ago, possess everything necessary to construct,
say, the N-soliton solutions for the DS equation.

\subsection{The Ablowitz-Ladik hierarchy and the KP equation.}

The result I want present in this section is, in some sense, the most
important example of 'embedding' into the ALH approach. To derive it one
has to consider the system of three first equations of the 'positive'
subhierarchy, that is the system consisting of the equations (\ref{1q}),
(\ref{2q}) and the equation determining the third flow,
$\partial / \partial z_{3}$:

\begin{eqnarray}
\partial_{1} q_{n} &=& -ip_{n}q_{n+1}
\\
\partial_{2} q_{n} &=&
 -ip_{n}p_{n+1}q_{n+2}+ir_{n-1}p_{n}q_{n}q_{n+1}+ip_{n}r_{n}q_{n+1}^{2}
\\
\partial_{3} q_{n} &=&
-ip_{n}p_{n+1}p_{n+2}q_{n+3}
+ip_{n}p_{n+1}r_{n+1}q_{n+2}^{2}
+2ip_{n}r_{n}p_{n+1}q_{n+1}q_{n+2}
\\&&
+ir_{n-1}p_{n}q_{n}p_{n+1}q_{n+2}
-ip_{n}r_{n}^{2}q_{n+1}^{3}
-ir_{n-1}p_{n}q_{n+1}^{2}
\cr&&
+2ir_{n-1}p_{n}^{2}q_{n+1}^{2}
-ir_{n-1}^{2}p_{n}q_{n}^{2}q_{n+1}
+ir_{n-2}p_{n-1}p_{n}q_{n}q_{n+1}
\nonumber
\end{eqnarray}
By very lengthy but straightforward calculations one can verify that for
all $n$'s the quantity

\begin{equation}
u = -\kappa \bar q_{n-1}p_{n}q_{n+1}
\end{equation}
satisfies the KP equation:

\begin{equation}
\partial_{1}
\left(
4 \partial_{3} u
+ \partial_{111} u
+ 12 u \, \partial_{1} u
\right) =
3 \partial_{22} u
\end{equation}
Thus, \magicphrase{the KP equation}.

\subsection{The Ablowitz-Ladik hierarchy and the AKNS hierarchy.}

It follows from equations (\ref{1q}) -- (\ref{2r}) that the quantities
$q_{n}$ and $r_{n-1}$ satisfy the closed system:

\begin{eqnarray}
&&
i\partial_{2}q_{n} +
\partial_{1}^{2}q_{n} + 2i q_{n}r_{n-1} \partial_{1}q_{n} = 0
\label{systDNLSE_1}
\\&&
i\partial_{2}r_{n-1} -
\partial_{1}^{2}r_{n-1} + 2i q_{n}r_{n-1} \partial_{1}r_{n-1} = 0
\label{systDNLSE_2}
\end{eqnarray}
This is one of the forms of the derivative NLSE. To rewrite it in the
standard way one can use the gauge transform which leads to the following
result: the quantities

\begin{equation}
Q = q_{n} \exp(-i\phi), \qquad
R = r_{n-1} \exp(i\phi)
\end{equation}
where

\begin{equation}
\phi = \int^{z_{1}} dz \, q_{n}(z,z_{2})r_{n-1}(z,z_{2})
\end{equation}
satisfy the system

\begin{eqnarray}
&&
i\partial_{2} Q +
\partial_{1}^{2} Q + 2i \partial_{1} Q^{2}R = 0
\\&&
i\partial_{2} R -
\partial_{1}^{2} R + 2i \partial_{1} QR^{2} = 0
\end{eqnarray}
which is the complexified derivative NLSE (I use the term
'complexified' to indicate that $Q$ and $R$ are not related by the
involution $Q=\pm\bar R$ which is typical for physical applications).
Thus, \magicphrase{the derivative NLSE}.

Having obtained this result, one can expect that similar relations exist
between the ALH and AKNS hierarchy because, using the paraphrase of
Kipling's words given by Newell, "The AKNS thing and the DNLS string are
sisters under the skin" \cite{N}.  However this cannot be done using only
local in $q_{n}$'s and $r_{n}$'s functions, but is possible in terms of
the $\tau$-functions of the ALH,

\begin{equation}
p_{n} = { \tau_{n-1} \tau_{n+1} \over \tau_{n}^{2} }
\end{equation}
(see (\ref{tau-def}) below). It follows from
(\ref{1q}) -- (\ref{2r}) that the quantities
$Q$ and $R$ given by

\begin{equation}
Q = { \tau_{n+1} \over \tau_{n} } q_{n+1},
\qquad
Q = { \tau_{n-1} \over \tau_{n} } r_{n-1}
\end{equation}
solve the complexified NLSE

\begin{eqnarray}
i\partial_{2} Q+\partial_{11} Q + 2 Q^{2}R = 0
\\
-i\partial_{2} R+\partial_{11} R + 2QR^{2} = 0
\end{eqnarray}
Moreover, considering also the third flow one can derive that $Q$ and $R$
solve the first of the higher DNLSE's as well:

\begin{eqnarray}
\partial_{3} Q + \partial_{111} Q + 6QR \partial_{1} Q = 0
\\
\partial_{3} R + \partial_{111} R + 6QR \partial_{1} R = 0
\end{eqnarray}
This is a demonstration of the fact that \magicphrase{the AKNS hierarchy}.

\section{'Embedding' into the 'vectorized' ALH.} \label{sec-gauged}
\setcounter{equation}{0}

The relations between the O(3,1) $\sigma$-model, 2DTL, DS, KP equations
and the ALH, discussed in the previous sections were formulated in terms
of $q_{n}$'s and $r_{n}$'s: {\it solutions} of the equations of the ALH
yield solutions to other problems. Now I want to present some results on
the 'embedding' not into the ALH itself but into a set of related
equations, which are equations for some combinations of the solutions of
the linear problems (\ref{zcr-sp}), (\ref{zcr-evol}) associated with the
ALH. Such relation is called in literature 'gauge equivalence':
NLSE is gauge equivalent to the Landau-Lifshitz equation, sine-Gordon ---
to the Thirring model, DS equation --- to the Ishimori spin model.

Consider the matrices $\sigma_{n}^{a}, a=1,2,3$ defined by

\begin{equation}
\sigma_{n}^{a} =
  \Psi_{n}^{-1} \sigma^{a} \Psi_{n}
\label{sigma}
\end{equation}
where $\sigma^{a} (a=1,2,3)$ are the Pauli matrices

\begin{equation}
\sigma^{1} = \pmatrix{0 & 1 \cr 1 & 0}, \qquad
\sigma^{2} = \pmatrix{0 & -i \cr i & 0}, \qquad
\sigma^{3} = \pmatrix{1 & 0 \cr 0 & -1}.
\label{pauli}
\end{equation}
and $\Psi_{n}$ is a $2 \times 2$ invertible matrix solution of the system
(\ref{zcr-sp}), (\ref{zcr-evol})

\begin{eqnarray}
\Psi_{n+1} &=& U_{n} \Psi_{n}
\label{ll:zcr-sp}\\
\partial_{j} \Psi_{n} &=& V_{n}^{j} \Psi_{n}
\label{ll:zcr-evol}
\end{eqnarray}
It follows from (\ref{sigma}) and (\ref{ll:zcr-evol}) that

\begin{equation}
\partial_{j} \sigma^{a}_{n} =
\Psi_{n}^{-1} \left[ \sigma^{a}, V_{n}^{j} \right]  \Psi_{n}
\label{sigma-evol}
\end{equation}
The right-hand sides of these equations can be written in terms of the
matrices $\sigma^{a}_{n}$ for different $a$'s and $n$'s.
Moreover, we can obtain the closed system of equations for the matrices
$\sigma^{a}_{n}$ with the index $a$ being fixed (in what follows we
will deal with the matrices $\sigma^{3}_{n}$) by means of the 'scattering'
problem (\ref{ll:zcr-sp}) which can be used to relate $\sigma^{a}_{n}$'s
and $\sigma^{3}_{n \pm 1}$

\begin{eqnarray}
p_{n} \sigma^{3}_{n+1} &=&
\left( 1 + q_{n}r_{n} \right) \sigma^{3}_{n} +
2 \lambda^{-1} r_{n} \sigma^{+}_{n} -
2 \lambda q_{n} \sigma^{-}_{n}
\label{sigma+1}
\\
p_{n-1} \sigma^{3}_{n-1} &=&
\left( 1 + q_{n-1}r_{n-1} \right) \sigma^{3}_{n} -
2 \lambda r_{n-1} \sigma^{+}_{n} +
2 \lambda^{-1} q_{n-1} \sigma^{-}_{n}
\label{sigma-1}
\end{eqnarray}
where the matrices $\sigma^{\pm}_{n}$ are given by

\begin{equation}
\sigma^{\pm}_{n}=
 {1 \over 2} \left(\sigma^{1}_{n} \pm i\sigma^{2}_{n}\right)
\label{sigma-pm}
\end{equation}
These identities also enable to express some combinations of the "ALH
quantities" $q_{n}$, $r_{n}$ and $p_{n}$ in terms of the "matrix" ones.
For example, it follows from (\ref{sigma+1}) that

\begin{equation}
p_{n} =
{ 2 \over 1 + {1 \over 2} \; {\rm tr} \; \sigma_{n+1} \sigma_{n} }
\label{p-sigma}
\end{equation}
For our further purposes it is convenient to use the matrix-vector
correspondence

\begin{equation}
S = \sum_{a=1}^{3} S_{a} \sigma^{a}
\; \longrightarrow \;
\vec S = \left( S_{1}, S_{2}, S_{3} \right)
\label{matr-vec}
\end{equation}
where $\sigma^{a}$ are the Pauli matrices (\ref{pauli}).

The equations (\ref{sigma-evol}) together with (\ref{sigma+1}),
(\ref{sigma-1}) can be viewed then as a system of DDE's for the matrices
$\sigma^{3}_{n}$ or the vectors $\vec\sigma_{n}$ which
correspond to $\sigma^{3}_{n}$'s by (\ref{matr-vec}),

\begin{equation}
\sigma^{3}_{n}
\; \longrightarrow \;
\vec \sigma_{n}
\end{equation}

Namely these equations will play the role of the starting system of DDE's,
and from them I will derive some interesting difference and differential
consequences.  Our first example is the 'gauge' analog of the O(3,1)
$\sigma$-model.

\subsection{The Ablowitz-Ladik hierarchy and the Landau-Lifshitz equation.}

Using the expressions for $V_{n}^{\pm 1}$,

\begin{equation}
V_{n}^{1} =
- i \pmatrix{
0 & \lambda^{-1} r_{n-1} \cr
\lambda^{-1} q_{n} & \lambda^{-2} - r_{n-1}q_{n} }
, \quad
V_{n}^{-1} =
i \pmatrix{
\lambda^{2} - q_{n-1}r_{n} & \lambda r_{n} \cr
\lambda q_{n-1} & 0  }
\end{equation}
one can express the derivatives of the matrices $\sigma^{3}_{n}$ in terms
of the matrices $\sigma^{\pm}_{n}$ given by (\ref{sigma-pm}) as follows:

\begin{eqnarray}
(i\lambda / 2)\; \partial \sigma^{3}_{n} &=&
  r_{n-1} \sigma^{+}_{n} - q_{n} \sigma^{-}_{n},
\label{dsdz}
\\
(i / 2\lambda)\; \bar\partial \sigma^{3}_{n} &=&
   - r_{n} \sigma^{+}_{n} + q_{n-1} \sigma^{-}_{n}.
\label{dsdbz}
\end{eqnarray}
(the symbols $\partial$ and $\bar\partial$ stand, remind,  for
$\partial_{1}$ and $\bar\partial_{1}$). From these relations, using
analogous expressions for the derivatives $\partial\sigma^{\pm}_{n}$,
$\bar\partial\sigma^{\pm}_{n}$ and formulae (\ref{1q}), (\ref{-1q}), one
can obtain, after straightforward calculations, omitted here, that for
every $n$ the matrix $S = \sigma^{3}_{n}$ solves the equation

\begin{equation}
\partial \bar\partial S +
{ 1 \over 2 } \;
\left({\rm tr} \; \partial S \, \bar\partial S \right) S +
{ 1 \over 2i }
\left[
  \lambda^{2} \partial S +
  \lambda^{-2} \bar\partial S\, , \,\
  S
\right]
= 0
\label{ll-matrix}
\end{equation}
In the vector form equation (\ref{ll-matrix}) becomes

\begin{equation}
\vec S_{z \bar z} +
\left( \vec S_{z} \vec S_{\bar z} \right) \vec S +
\left[
  \left(
  \lambda^{2} \vec S_{z} + \lambda^{-2} \vec S_{\bar z}
  \right)
\; \times \; \vec S
\right] = 0
\label{ll-vec}
\end{equation}
which turns out to be the (0+2)-dimensional version of the
Landau-Lifshitz equation for the isotropic two-dimensional
classical Heisenberg ferromagnets,

\begin{equation}
\partial_{t}\vec S = g \left[ \vec S \times \Delta\vec S \right],
\qquad
\vec S^{2} = 1
\label{heis}
\end{equation}
where  $\Delta$ is the two-dimensional Laplacian.
Indeed, if we restrict ourselves with the stationary structures of the
form $ \vec S = \vec S \left( x - v_{x}t, y - v_{y}t \right) $, then
equation (\ref{heis}) becomes

\begin{equation}
g \left[ \vec S \times \Delta\vec S \right] +
\left( \vec v \, , \, \nabla \vec S \right) = 0.
\end{equation}
Introducing the variables

\begin{equation}
z = {v \over 4g}
\left[ x - v_{x}t + i( y - v_{y}t) \right],
\quad
\bar z = {v \over 4g}
\left[ x - v_{x}t - i( y - v_{y}t) \right]
\end{equation}
where $v = \left| \vec v \right|$, the latter can be rewritten as

\begin{equation}
\left[ \vec S \times \vec S_{z \bar z} \right] +
\lambda^{2} \vec S_{z} + \lambda^{-2} \vec S_{\bar z} = 0
\label{Tijon}
\end{equation}
with $\lambda=\exp(i\gamma/2)$, where the angle $\gamma$ is defined by
$v_{x}=v\cos\gamma$, $v_{y}=v\sin\gamma$ which is equivalent to
(\ref{ll-vec}).

Thus,
\magicphrase{the (0+2)-dimensional Landau-Lifshitz equation}.
The equation (\ref{ll-vec}) is known to be integrable (its zero-curvature
representation one can find in the paper \cite{Papa1}), and one can
tackle it by elaborating the corresponding inverse scattering transform,
but, as in the cases discussed above, we can now obtain a wide range of
physically interesting solutions using the already known solutions for the
ALH. This has been done in the paper \cite{ll}.

\subsection{
The Ablowitz-Ladik hierarchy and a multi-$\sigma$-field model.}
In this section we will fix the quantity $\lambda$,

\begin{equation}
\lambda = 1
\end{equation}
noting that the case of the arbitrary $\lambda$'s ($|\lambda|=1$) can be
restored by the rotation of the coordinates $x$ and $y$ ($z \equiv z_{1} =
x + i y$).

Differentiating (\ref{dsdz}) with respect to $\bar z$ (or
(\ref{dsdbz}) with respect to $z$, which leads to the same result) one
can obtain

\begin{equation}
\partial\bar\partial \sigma_{n} =
   2 \left( 2 - p_{n-1} - p_{n} \right)\sigma_{n} +
   2 \left( r_{n} - r_{n-1}  \right) \sigma^{+}_{n} +
   2 \left( q_{n-1} - q_{n}  \right) \sigma^{-}_{n}
\label{dds}
\end{equation}
Using the identities

\begin{equation}
2 \left( r_{n} - r_{n-1}  \right) \sigma^{+}_{n} +
2 \left( q_{n-1} - q_{n}  \right) \sigma^{-}_{n} =
p_{n} \sigma_{n+1} +
\left( p_{n-1} + p_{n} - 4 \right)\sigma_{n} +
p_{n-1} \sigma_{n-1}
\end{equation}
which follow from (\ref{sigma+1}), (\ref{sigma-1}) and (\ref{p-sigma})
one can rewrite (\ref{dds}) as

\begin{equation}
\partial\bar\partial \sigma_{n} =
   p_{n-1} \sigma_{n-1} -
   \left( p_{n-1} + p_{n} \right) \sigma_{n} +
   p_{n} \sigma_{n+1}
\end{equation}
or, using again (\ref{dsdz}) - (\ref{dsdbz}), as

\begin{equation}
\partial \bar\partial \sigma^{3}_{n} +
{ 1 \over 2 }
   \left( {\rm tr}\;
   \partial \sigma^{3}_{n} \bar\partial \sigma^{3}_{n}
   \right)
   \sigma^{3}_{n} =
   p_{n-1} \sigma_{n-1} +
   \left( p_{n-1} + p_{n} - 4 \right) \sigma_{n} +
   p_{n} \sigma_{n+1}
\label{ms-matrix}
\end{equation}
The last equation turns out to be the field equation of some model which
to author's knowledge is new and which can be viewed as a multi-field
generalization of the well-known O(3) $\sigma$-model.

The O(3) $\sigma$ model is described by the Lagrangian

\begin{equation}
{\cal L} = {\cal T} + {\cal L}^{'}
\end{equation}
where ${\cal T}$ is the 'kinetic-energy' term in two dimensions

\begin{equation}
{\cal T} =
{ 1 \over 2 }
\left( \nabla \vec S , \nabla \vec S \right),
\qquad
\nabla = \left( \partial_{x}, \partial_{y} \right)
\label{kinetic-one}
\end{equation}
and ${\cal L}^{'}$ takes into account the restriction

\begin{equation}
{\vec S}^{2} = 1
\end{equation}
Our generalization is, first, in taking instead of one field $\vec S$ an
infinite number of fields

\begin{equation}
\vec S_{n},
\qquad
n = 0, \pm 1, ...,
\qquad
{\vec S_{n}}^{2} = 1
\end{equation}
i.e., in replacing ${\cal T}$ (\ref{kinetic-one}) with

\begin{equation}
{\cal T} =
{ 1 \over 2 }
\sum_{n}
\left( \nabla \vec S_{n} , \nabla \vec S_{n} \right)
\end{equation}
It is easily understood that in absence of interactions between different
fields (spins) our generalization would be trivial and of little
interest both from physical and mathematical standpoints. So, we consider
also the interaction, the nearest-neighbour one,

\begin{equation}
{\cal U} =
\sum_{n}
u\left( \vec S_{n} , \vec S_{n+1} \right)
\end{equation}
In what follows our attention will be restricted to the particular pair
potential $u$ which has no, so to say, physical origin, but is remarkable
by the fact that it preserves integrability of the corresponding
equations.  Namely, we will study the potential of the classical
Heisenberg ferromagnets model (which is known to be integrable,
see, e.g., \cite{FT,Papa2} and to be in close relations with the ALH
\cite{Ishim2}):

\begin{equation}
{\cal U} =
- J
\sum_{n}
\ln \left[
1 + \left( \vec S_{n} , \vec S_{n+1} \right)
\right]
\label{potential}
\end{equation}
Summarizing, the model considered is given by

\begin{equation}
{\cal L} =
{ 1 \over 2 }
\sum_{n}
\left( \nabla \vec S_{n} , \nabla \vec S_{n} \right)
+
J \sum_{n}
\ln \left[
1 + \left( \vec S_{n} , \vec S_{n+1} \right)
\right]
+
{\cal L}^{'}
\label{lagrangian}
\end{equation}
where

\begin{equation}
{\cal L}^{'} =
\sum_{n}
\Lambda_{n}
\left( {\vec S_{n}}^{2} - 1 \right)
\end{equation}
and $\Lambda_{n}$'s are the Lagrange multipliers, which should be
chosen to satisfy the conditions  ${\vec S_{n}}^{2}=1$.

The field equations corresponding to (\ref{lagrangian}) are of the form

\begin{equation}
\Delta \vec S_{n} = \vec {\cal F}_{n} + \Lambda_{n} \vec S_{n}
\end{equation}
where

\begin{equation}
\vec{\cal F}_{n} =
- { \partial \over \partial\vec S_{n} } {\cal U} =
{ 1 \over 2} J
\left( f_{n-1} \vec S_{n-1} + f_{n} \vec S_{n} \right)
\end{equation}
with

\begin{equation}
f_{n} =
{ 2
  \over
  1 + \left( \vec S_{n}, \vec S_{n+1} \right)
}
\label{f-sigma}
\end{equation}
and after calculating $\Lambda_{n}$'s they become

\begin{equation}
\Delta \vec S_{n} +
\left( \nabla\vec S_{n}, \nabla\vec S_{n} \right)\vec S_{n} =
{ 1 \over 2} J
\left\{
   f_{n-1} \vec S_{n-1} +
   \left( f_{n-1} + f_{n} - 4 \right) \vec S_{n} +
   f_{n} \vec S_{n+1}
\right\}
\label{ms-vector}
\end{equation}
These equations after identifying $f_{n}$ and $p_{n}$ (compare
(\ref{f-sigma}) and (\ref{p-sigma})) and
rescaling the coordinates become the vector form of
the matrix equations (\ref{ms-matrix}). Thus, {\it the
multi-$\sigma$-field model} described above, which is a (2+1)-dimensional
system gauge equivalent to the 2DTL, \magicphrase{}.

\subsection{The Ablowitz-Ladik hierarchy and the Ishimori equation.}

The following example is a manifestation of the already known fact that
the Ishimori model is gauge equivalent to the DS equation \cite{LS2}.

Using the expressions for $V_{n}^{(\pm 1)}$, $V_{n}^{(\pm 2)}$ one can
express the derivatives of the matrices $\sigma^{3}_{n}$ as follows:

\begin{eqnarray}
(i\lambda / 2)\; \partial_{1} \sigma^{3}_{n} &=&
\phantom{-}
  r_{n-1} \sigma^{+}_{n} - q_{n} \sigma^{-}_{n}, \\
(i / 2\lambda)\; \bar\partial_{1} \sigma^{3}_{n} &=&
   - r_{n} \sigma^{+}_{n} + q_{n-1} \sigma^{-}_{n}, \\
\cr
(i\lambda / 2)\; \partial_{2} \sigma^{3}_{n} &=&
\phantom{-}
   \left( \lambda^{-2} r_{n-1} + r_{n-2}p_{n-1} - r_{n-1}^{2}q_{n} \right)
   \sigma^{+}_{n}
\cr&&
   - \left( \lambda^{-2} q_{n} + p_{n}q_{n+1} - r_{n-1}q_{n}^{2} \right)
   \sigma^{-}_{n},
\\
(i / 2\lambda)\; \bar\partial_{2} \sigma^{3}_{n} &=&
   - \left( \lambda^{2} r_{n} + p_{n}r_{n+1} - q_{n-1}r_{n}^{2} \right)
   \sigma^{+}_{n}
\cr&&
   + \left( \lambda^{2}q_{n-1} + q_{n-2}p_{n-1} - q_{n-1}^{2}r_{n} \right)
   \sigma^{-}_{n}.
\end{eqnarray}

From these relations, using analogous expressions for the derivatives
$\partial_{j}\sigma^{\pm}_{n}$, one can obtain that for every $n$ the
matrix $S=\sigma^{3}_{n}$ (the index $n$ will be fixed and omitted
hereafter) with $|\lambda|=1$ satisfies the following equations:

\begin{eqnarray}
- 2i \partial_{2} S + \left[ S , \partial^{2} S \right] &=&
4 i w \partial_{1} S,
\label{ishim-p}
\\
- 2i \bar\partial_{2} S + \left[ S , \bar\partial^{2} S \right] &=&
4 i \bar w \bar\partial_{1} S,
\label{ishim-n}
\end{eqnarray}
where

\begin{equation}
w \equiv r_{n-1}q_{n} - \lambda^{-2}
, \quad
\overline w \equiv q_{n-1}r_{n} - \lambda^{2}.
\end{equation}
From the other hand, it follows from (\ref{1q})-(\ref{1r}) that

\begin{equation}
i \bar\partial_{1} w = - i \partial_{1} \overline w =
p_{n} - p_{n-1}.
\label{dw}
\end{equation}
The r.h.s. of (\ref{dw}) can be expressed in terms of the matrix $S$

\begin{equation}
p_{n} - p_{n-1} =
- {1 \over 8} {\rm tr}\,
\left\{ S \left[ \partial_{1}S \, , \, \bar\partial_{1}S \right] \right\}.
\end{equation}
Finally, using the real variables $x,y,t$ (\ref{xyt}) and the
quantity $f$ defined by

\begin{equation}
w = - i \partial f
\end{equation}
which satisfy the equation
$\Delta f = 4 \left( p_{n} - p_{n-1} \right)$,
equations (\ref{ishim-p}), (\ref{ishim-n}) and (\ref{dw}) can be presented
as

\begin{eqnarray}
i  S_{t} &=&
{1 \over 4} \left[ S , \Box S \right] +
i \left( f_{y} S_{x} + f_{x} S_{y} \right),
\label{I1}\\
\Delta f &=&
{1 \over 4i} {\rm tr}\,
\left\{ S \left[ S_{x} , S_{y} \right] \right\}.
\label{I2}
\end{eqnarray}
This system, when rewritten in terms of the vector $\vec S$ related to the
matrix $S$ by (\ref{matr-vec}) becomes the Ishimori equation \cite{Ishim}

\begin{eqnarray}
\vec S_{t} &=&
{1 \over 2} \left[ \vec S \times \Box \vec S \right] +
f_{y} \vec S_{x} + f_{x} \vec S_{y},
\label{Ishim1}\\
\Delta f &=&
\left( \vec S \cdot \left[ \vec S_{x} \times \vec S_{y} \right] \right).
\label{Ishim2}
\end{eqnarray}
This is the main result of this section:
\magicphrase{the Ishimori equation}.

\section{Functional representation of the Ablowitz-Ladik hierarchy.}
\label{sec-newell}
\setcounter{equation}{0}

In the above sections one can find many examples of the relations
between the ALH and other integrable models. These examples show that the
ALH possesses some kind of 'universality'. A hypothesis arises that all
known integrable models can be 'embedded' into the ALH. So, the main
question one should answer now is the following: is the ALH in some sense
a distinguished hierarchy?  It seems to me that there is some sense in
these words.  I cannot give any rigorous proof of such a statement (and
even formulate it in strict terms). Nevertheless, I think this question
is worth studying. The results discussed above are in some sense
'empirical' facts: one can easily verify them by simple calculations, but
one can hardly find there an answer to the question why do such apparently
different models turn out to be interrelated. To do this one probably has
to consider the problem, and hence the ALH, in some more general terms.
Namely this was the motivation of the work \cite{newell}.

The main result of the paper \cite{newell} is the representation
of the ALH, which has been originally introduced as an infinite system of
differential-difference equations, as a finite system of
difference-functional equations. I will not repeat here the calculations
outlined there and only write down the main result. The 'positive'
Ablowitz-Ladik subhierarchy is shown to be equivalent to following
equations:

\begin{eqnarray}
&&
\sigma_{n}  \left( z + \d \right) \tau_{n}  \left( z - \d \right) -
\tau_{n}    \left( z + \d \right) \sigma_{n}\left( z - \d \right)
=
\label{main-q}
\\&&
\hspace{3cm} =
\delta \;
\sigma_{n+1}\left( z + \d \right) \tau_{n-1}\left( z - \d \right)
\cr&&
\tau_{n}  \left( z + \d \right) \rho_{n}  \left( z - \d \right) -
\rho_{n}  \left( z + \d \right) \tau_{n}  \left( z - \d \right) =
\label{main-r}
\\&&
\hspace{3cm} =
\delta \;
\tau_{n+1}\left( z + \d \right) \rho_{n-1}\left( z - \d \right)
\cr&&
\tau_{n}  \left( z + \d \right) \tau_{n}  \left( z - \d \right) -
\sigma_{n}\left( z + \d \right) \rho_{n}  \left( z - \d \right) =
\label{main-p}
\\&&
\hspace{3cm} =
\tau_{n+1}\left( z + \d \right) \tau_{n-1}\left( z - \d \right)
\nonumber
\end{eqnarray}
Here $z$ stands for $\left(z_{1}, z_{2}, ... , z_{k}, ... \right)$,
$[\delta]$ is the traditional now designation for
$\left(\delta, \delta^{2}/2, ... , \delta^{k}/k, ... \right)$,
and $\tau_{n}$, $\sigma_{n}$ and $\rho_{n}$ are $\tau$-functions
defined by

\begin{equation}
q_{n} = { \sigma_{n} \over \tau_{n} },
\qquad
r_{n} = { \rho_{n} \over \tau_{n} },
\qquad
p_{n} = { \tau_{n-1} \tau_{n+1} \over \tau_{n}^{2} }
\label{tau-def}
\end{equation}
Expanding (\ref{main-q})-(\ref{main-p}) in the power series in $\delta$
one can obtain all equations of the 'positive' subhierarchy.
As is seen, in the above formulae only the 'positive' coordinates $z_{j}$
are used. An analogous representation, involving $\bar z_{j}$'s, can be
obtained for the 'negative' subhierarchy.

Using Hirota's bilinear approach one can rewrite
(\ref{main-q})-(\ref{main-p}) as

\begin{equation}
\exp\left[ {i \over 2} D(\delta) \right]
\pmatrix{
  \sigma_{n} \cdot \tau_{n} -
  \tau_{n} \cdot \sigma_{n} -
  \delta \; \sigma_{n+1} \cdot \tau_{n-1}
\cr
  \rho_{n} \cdot \tau_{n} -
  \tau_{n} \cdot \rho_{n} +
  \delta \; \tau_{n+1} \cdot \rho_{n-1}
\cr
  \sigma_{n} \cdot \rho_{n} -
  \tau_{n} \cdot \tau_{n} +
  \tau_{n+1} \cdot \tau_{n-1}
} = 0
\label{Hirota-n}
\end{equation}
where

\begin{equation}
D(\delta) =
\sum_{k=1}^{\infty} { \delta^{k} \over k } D_{k},
\qquad
D_{j} = D_{\textstyle z_{j}}
\end{equation}
and the operators $D_{\textstyle z_{j}}$ are defined by

\begin{equation}
D_{x}^{a} ... D_{y}^{b} \; u \cdot v =
\left.
\left(
  {\partial \over \partial x} - {\partial \over \partial x^{'}}
\right)^{a}
...
\left(
  {\partial \over \partial y} - {\partial \over \partial y^{'}}
\right)^{b}
u(x, y, ...) v(x^{'}, y^{'}, ... )
\right|_{x^{'}=x, \, y^{'}=y, ... }
\end{equation}
Thus, we have a system of difference-functional equations  for an infinite
number of triplets of $\tau$-functions. It has been shown in \cite{newell}
that from this system one can derive a closed system for only one triplet
$\left( \sigma, \rho, \tau \right)$ :

\begin{equation}
\widehat G(\delta)
\pmatrix{
  \sigma \cdot \tau \cr
  \tau   \cdot \rho \cr
  \sigma \cdot \rho + \tau \cdot \tau} = 0
\label{pos-hie}
\end{equation}
where $\sigma$, $\rho$ and $\tau$ stand for $\sigma_{n}$, $\rho_{n}$ and
$\tau_{n}$ with $n$ being fixed and the operator $\widehat G(\delta)$ is
defined by

\begin{equation}
\widehat G(\delta) =
2 \sin\left[ {1 \over 2} D(\delta) \right] -
\delta \; D_{1} \;
\exp\left[ {i \over 2} D(\delta) \right]
\end{equation}
These equations, when expanded in power series in $\delta$ yield a
hierarchy of partial differential equations of the NLSE type.

This procedure of 'shortening' can be continued by excluding $\sigma$ and
$\rho$ and obtaining the 'scalar' equations for only one function, $\tau$,

\begin{equation}
\widehat H(\delta) \; \tau \cdot \tau = 0
\label{kp-hie}
\end{equation}
where the operator $\widehat H(\delta)$ is given by

\begin{equation}
\widehat H(\delta) =
\left[ 2 D_{1} - \delta \left(D_{2} + i D_{11} \right) \right]
\exp\left[ {i \over 2} D(\delta) \right]
\end{equation}

Expanding the operator $\widehat H(\delta)$ in
powers of $\delta$ one can obtain an infinite number of operators.
The first of them is

\begin{equation}
4 D_{31} - 3 D_{22} + D_{1111}
\end{equation}
i.e. the operator which appears in the bilinear representation of the KP
equation.

Thus, we have a chain: the infinite system  (\ref{main-q})-(\ref{main-p})
$\rightarrow$ the system of three equations (\ref{pos-hie})
$\rightarrow$ one equation (\ref{kp-hie}). This chain leads from the ALH,
through the NLSE-like hierarchies, to the KP-like one and may be useful to
understand the place of the ALH among other integrable hierarchies.

\section{Conclusion.}  \label{sec-conclusion}
\setcounter{equation}{0}

The results presented in this paper are aimed to demonstrate that the ALH
indeed possesses some kind of 'universality'. One more indication of, in
some sense, distinguished character of the ALH, which has not been
mentioned above and which will be discussed in a separate paper, comes
from the theory of the $\theta$-functions. It is well known that the
$\theta$-functions of the finite genus Riemann surfaces satisfy some
functional relations, I mean the Fay's trisecant identity \cite{Fay}.
There is also a well-elaborated procedure how to derive from the latter a
number of differential relations satisfied by the $\theta$-functions, and
one can easily find in literature a lot of identities involving some {\it
combinations} of some differential operators $\partial_{j}$ (usually ones
that appear in the KdV, KP, 2DTL and other equations). It is
surprising, but to author's knowledge the following, rather natural,
question has not been given due attention. Consider the problem of how to
express the action of the operators $\partial_{j}$ taken {\it separately}
on the $\theta$-functions (or some ratios of the $\theta$-functions,
$\theta(\zeta+\alpha)/\theta(\zeta+\beta)$) without invoking other
operators $\partial_{k}, (k \ne j)$, but only in terms of
$\theta$-functions, maybe of some other arguments. This problem is not
very difficult: it can be done using the already known (and widely used)
algorithm, by expanding the Fay's identity. It turns out that these
relations can be written in the form of the equations of the ALH. This
fact becomes more transparent if we consider this question in the
framework of the functional representation of the ALH: equations
(\ref{main-q}) -- (\ref{main-p}) in the quasiperiodical case become
equivalent to the Fay's identity.

However the word 'universality' often used in this paper should not be
understood literally (that is why I used it in the quotation marks)
because of the following.  The relations between the ALH and other
equations are not, in general, ones of equivalence.  The correspondence
between the ALH subsystem (\ref{sigma-system}) and the field equations of
the O(3,1) $\sigma$-model (\ref{nonlin-Helmholtz}) is one-to-one: one can
find in \cite{sigma} the inverse transform (from the field equations to
the DNLSE-DMKdV chain).  At the same time in most of the other examples
the situation is different.  Not all solutions of, say, (2+1)-dimensional
models can be obtained by the 'embedding' into the ALH method. And a
logical continuation of the works \cite{sigma,2dtl,ds,newell} is to study
the question of what kind of reduction do we have when 'split' some system
(2DTL, DS, KP) into few equations from the ALH. In the case of, e.g., the
DS equation this is easy to do: we can start from the fact that the
quantities $q_{n}$ satisfy not only the (2+1)-dimensional DS system
(\ref{dsq}), (\ref{dsa}) but also the 2-dimensional O(3,1) $\sigma$-model
equations, and to formulate the ansatz we have implicitly used. However in
other cases, the 2DTL for example, this problem is somewhat more
difficult: I cannot at present write down some, so to say, more simple
equations which are satisfied by the quantities $p_{n}$, i.e. I cannot
formulate, using the language of equations, what restrictions do we impose
on solutions of the 2DTL when substituting the latter by the ALH equations
(\ref{nonlin-Helmholtz}). I think some progress can be made in the
framework of a more general approach, e.g., as one discussed in
\cite{newell}.  Maybe it will be easier to understand the ALH-(other
equations) reductions studying the chain from the ALH to the KP equation
mentioned at the end of the previous section. To conclude, I want to
stress once more that this question is not answered yet, while it seems to
be one of the most important problems to be solved before (if ever) to
call the ALH universal (without the quotation marks).

\end{document}